\documentclass[aps,twocolumn,superscriptaddress,prl,amsmath]{revtex4-2}
\usepackage{multirow}
\usepackage[latin9]{inputenc}
\usepackage{mathrsfs}
\usepackage{txfonts}
\usepackage{amssymb}
\usepackage{graphicx,subfigure,float}
\usepackage{epstopdf}
\usepackage{dcolumn}
\usepackage{bbm}
\usepackage{bm}
\usepackage{color}
\usepackage[colorlinks, linkcolor=blue, citecolor=blue, urlcolor=blue]{hyperref}
\usepackage{lipsum}
\usepackage{gensymb}
\usepackage{pifont}

\begin{document}
\title{Symmetry Classification of Non-Relativistic Hidden Spin Polarization in Noncollinear Magnets}
\author{Yuzhong Hu}
\thanks{These authors contributed equally to this work.}
\affiliation{Key Laboratory of Low Dimensional Materials and Application Technology of Ministry of Education, School of Materials Science and Engineering, Xiangtan University, Xiangtan 411105, China}
\affiliation{Hunan Provincial Key laboratory of Thin Film Materials and Devices, School of Materials Science and Engineering, Xiangtan University, Xiangtan 411105, China}
\author{Pan Zhou$^{*}$}
\email{zhoupan71234@xtu.edu.cn}
\affiliation{Key Laboratory of Low Dimensional Materials and Application Technology of Ministry of Education, School of Materials Science and Engineering, Xiangtan University, Xiangtan 411105, China}
\author{Baoru Pan}
\affiliation{Key Laboratory of Low Dimensional Materials and Application Technology of Ministry of Education, School of Materials Science and Engineering, Xiangtan University, Xiangtan 411105, China}
\author{PengBo Lyu}
\affiliation{Hunan Provincial Key laboratory of Thin Film Materials and Devices, School of Materials Science and Engineering, Xiangtan University, Xiangtan 411105, China}
\author{Lizhong Sun}
\email{lzsun@xtu.edu.cn}
\affiliation{Key Laboratory of Low Dimensional Materials and Application Technology of Ministry of Education, School of Materials Science and Engineering, Xiangtan University, Xiangtan 411105, China}
\affiliation{Hunan Provincial Key laboratory of Thin Film Materials and Devices, School of Materials Science and Engineering, Xiangtan University, Xiangtan 411105, China}
\date{\today}
\begin{abstract}
Hidden spin polarization (HSP), in which spin-polarized states exist locally while the total spin polarization are hidden in momentum space, has been extensively studied in nonmagnetic and collinear magnetic systems, but remains largely unexplored in noncollinear magnets. Here we establish a unified symmetry framework for HSP in noncollinear magnetic materials based on spin-group theory. We show that spin symmetries systematically constrain nonrelativistic spin polarization, giving rise to four distinct split spin-texture (SST) types for each local sector, denoted as SST-1, SST-2, SST-3, and SST-4. Based on these splitting forms, together with the dimensionality of the associated local spin textures and the symmetry relations between different local sectors, we further classify HSP into three categories: HSP-1, HSP-2, and HSP-3. We illustrate these categories using tight-binding models and representative material examples, including SrFe$_2$Se$_2$O, USb, Sr$_2$Mn$_3$Sb$_2$O$_2$, PrFeAsO, and GdMn$_2$Si$_2$. A survey of the MAGNDATA database further identifies 133, 7, and 139 candidate noncollinear magnetic materials hosting HSP-1, HSP-2, and HSP-3, respectively. In addition, our symmetry analysis and first-principles calculation show that many of these materials can exhibit nonzero spin-related response tensors. These results establish a general framework for understanding HSP in noncollinear magnets and highlight their potential for spin-dependent functionalities.\\
\end{abstract}
\maketitle
\indent \emph{Introduction.}---Spintronics has conventionally relied on materials with robust spin-polarized electronic states\cite{spin1,spin2,spin3,spin4,spin5,spin6, spin7,spin8}, in which spin polarization arises from exchange fields or from spin-orbit coupling (SOC) in noncentrosymmetric crystals\cite{SOC_1,SOC_2}. Beyond this paradigm, recent studies have established that even centrosymmetric crystals may host hidden spin polarization (HSP), featuring locally spin-polarized states in real space together with a momentum-dependent but globally compensated spin texture in reciprocal space\cite{hsp1,hsp2,hsp3,hsp4,hsp5,hsp6,hsp7,hsp8,hsp9,hsp10,hsp11,hsp12,hsp13,hsp14,hsp15,hsp16, hsp17}. Initially explored in nonmagnetic systems, HSP has more recently been extended to collinear antiferromagnets\cite{afm_hsp1,afm_hsp2,afm_hsp3, afm_hsp4} and linked to a variety of spin-dependent transport\cite{spin_transport1,spin_transport2} and optical responses\cite{optical1,optical2, optical3}, including spin-orbit torque\cite{SOT_1,SOT_2} and nonlinear transport effects\cite{nahc1,nahc2,nahc3}. These results considerably expand the scope of materials relevant to spintronics.\\
\indent In contrast to collinear magnets, noncollinear magnetic materials offer a far richer symmetry framework for unconventional spin polarization and have thus attracted considerable attention\cite{NAFM_1,NAFM_2,NAFM_3,NAFM_4,NAFM_5,NAFM_6,NAFM_7,Odd_1,Odd_2,Odd_3,Odd_4,Odd_5,Odd_6,Odd_7}. The intricate interplay between crystal symmetry and noncollinear magnetic symmetry enables spin-dependent band structures and transport responses beyond the conventional collinear paradigm\cite{transport1,transport2,transport3, transport4,transport5,transport6,transport7,transport8,transport9,transport10}. Despite the growing interest in the electronic, magnetic, and transport properties of noncollinear systems, their HSP remains largely unexplored. Moreover, the criteria established for collinear antiferromagnets presuppose a global spin-quantization axis and break down for noncollinear order\cite{afm_hsp2}, calling for a full spin-space-group treatment. In particular, a systematic understanding of HSP in noncollinear magnets, and of its relation to symmetry and observable responses, is still missing.\\
\indent Motivated by this gap, in this Letter we establish a unified symmetry framework for HSP in noncollinear magnets based on spin-group theory. Spin symmetries systematically constrain the nonrelativistic spin polarization of each local sector to four distinct split spin-texture (SST) types, SST-1--SST-4. Combining these with the dimensionality of the corresponding local spin textures and the symmetry relations between local sectors classifies HSP into three categories, HSP-1--HSP-3. We demonstrate them in tight-binding models and representative materials, identify numerous candidates through a survey of the MAGNDATA database, and show that many HSP-hosting noncollinear magnets allow nonzero spin-related response tensors, linking hidden spin textures to observable spin-dependent phenomena.\\
\indent \emph{Spin-texture prototypes in noncollinear magnets.}---The spin texture is defined as
$\mathbf{S}(\mathbf{k})=\langle\psi_{\mathbf{k}}|\boldsymbol{\sigma}|\psi_{\mathbf{k}}\rangle$ for each $\mathbf{k}$ in momentum space,
where $\boldsymbol{\sigma}$ denotes the Pauli matrices and $|\psi_{\mathbf{k}}\rangle$ the corresponding Bloch state. The HSP arises when a crystal can be decomposed into two real-space sectors, denoted as $\alpha$ and $\beta$, whose sector-resolved electronic states develop spin-polarized textures in momentum space. The two sectors are connected by a symmetry operation that enforces band degeneracy at every crystal momentum $\mathbf{k}$ and, at the same time, requires their spin textures to be opposite, namely
\[
\mathbf{S}^{\alpha}(\mathbf{k}) = -\mathbf{S}^{\beta}(\mathbf{k}),
\]
even though the electronic states in each sector still retain a finite local spin polarization. Therefore, to identify possible realizations of HSP in noncollinear magnets, one must first classify all symmetry-allowed nonrelativistic spin textures.

In the nonrelativistic limit, spin rotations are independent of spatial operations. The symmetry of a magnetic crystal is therefore naturally described by the spin space group (SSG)\cite{sgroup1,sgroup2,sgroup3,sgroup4}, which can generally be written as
\begin{equation}
\mathbf{G}_{SS}=\mathbf{G}_{SO}\times \mathbf{G}_{NS},
\end{equation}
where $\mathbf{G}_{SO}$ denotes the spin-only subgroup and $\mathbf{G}_{NS}$ contains operations acting jointly on spin and real space. For noncoplanar magnetic order $\mathbf{G}_{SO}=\{E\}$, while for coplanar order $\mathbf{G}_{SO}=\{E,TU_{\bm n}(\pi)\}$. The nontrivial symmetry elements in $\mathbf{G}_{NS}$ take the form
$g=[U_s \Vert R_g \vert \tau]$,
where $\{R_g|\tau\}$ acts on spatial coordinates, with $R_g$ belonging to $\{C_n(\theta),PC_n(\theta)\}$ and $\tau$ a fractional translation, while $U_s\in\{U_n(\phi),TU_n(\phi)\}$ represents spin rotations with the potential antiunitary time-reversal operation $T$.

\indent Because noncollinear magnetic order does not select a single global spin-quantization axis, the momentum-space spin polarization allowed in such systems is generally much richer than that in collinear magnets. For a spin texture $\mathbf{S}(\mathbf{k})$, invariance under a SSG operation $g=[U_s \Vert R_g \vert \tau]$ requires
\begin{equation}\label{eq2}
\mathbf{S}(\mathbf{k}) = U_s\, \mathbf{S}(\eta_g R_g^{-1}\mathbf{k}),
\end{equation}
where $\eta_g \equiv \det|U_s| = +1$ ($-1$) for unitary (antiunitary) operations in spin space. Thus, the symmetry constraint is fully determined by the transformed momentum $\tilde{\mathbf{k}}=\eta_g R_g^{-1}\mathbf{k}$. For a generic $\mathbf{k}$, the relevant SSG operations can be divided into three classes. The first class, denoted by $g_a$, satisfies $\tilde{\mathbf{k}}=\mathbf{k}$. These operations constrain the spin vector at the same momentum and therefore determine the dimensionality $d_s$ of the allowed spin texture, i.e., whether it is collinear ($d_s=1$), coplanar ($d_s=2$), or noncoplanar ($d_s=3$). After a suitable choice of spin axes, typical representatives of this class can be written as $[U_z(\phi)\Vert E\vert \tau]$ or $[TC_{2z}\Vert P]$. The second class, $g_b$, satisfies $\tilde{\mathbf{k}}=-\mathbf{k}$. These operations connect opposite momenta and determine the parity of each symmetry-allowed spin component under $\mathbf{k}\rightarrow -\mathbf{k}$: depending on the action of $U_s$, a component is constrained to be either even or odd. The third class, $g_c$, satisfies $\tilde{\mathbf{k}}\neq \pm\mathbf{k}$. Such operations relate spin textures at distinct symmetry-equivalent momenta in the Brillouin zone. They are important for constructing the full momentum-space texture, but do not by themselves fix either $d_s$ or the parity under $\mathbf{k}\rightarrow -\mathbf{k}$.

\indent The local spin texture is therefore specified by two independent pieces of information: the allowed spin subspace determined by $g_a$, and the parity of the allowed components determined by $g_b$. Their interplay gives rise to four generic SST types, summarized as follows.\\
\indent (i) SST-1:
SST-1 occurs when the symmetry constraints are provided only by $g_a$-type operations, so that symmetry fixes the dimensionality of the spin texture but imposes no definite relation between $\mathbf{S}(\mathbf{k})$ and $\mathbf{S}(-\mathbf{k})$. Accordingly, SST-1 can be collinear, coplanar, or noncoplanar, with the detailed symmetry conditions listed in Tab.~S1. A representative tight-binding model is given in Fig.~S1.\\
\indent (ii) SST-2:
SST-2 occurs when both $g_a$- and $g_b$-type symmetries are present, and the relevant $g_b$ operations act trivially on all symmetry-allowed spin components. In this case, every nonvanishing component satisfies $S_i(\mathbf{k})=S_i(-\mathbf{k})$, so the spin texture is even in momentum. As in SST-1, the dimensionality may still be collinear, coplanar, or noncoplanar, depending on the $g_a$-type constraints. The corresponding symmetry conditions are summarized in Tab.~S2, and a $d_s=1$ tight-binding example is shown in Fig.~S2.\\
\indent (iii) SST-3:
SST-3 also requires both $g_a$- and $g_b$-type symmetries, but now the relevant $g_b$ operations reverse all symmetry-allowed spin components, leading to $S_i(\mathbf{k})=-S_i(-\mathbf{k})$ for every nonvanishing component. The resulting texture is therefore odd in momentum. As a consequence, the Brillouin-zone-integrated spin polarization vanishes identically, irrespective of whether $g_c$-type symmetries are present. SST-3 can again be collinear, coplanar, or noncoplanar, with the symmetry conditions listed in Tab.~S3 and a $d_s=1$ model shown in Fig.~S3.\\
\indent (iv) SST-4:
SST-4 arises when $g_a$-type symmetries leave a multidimensional spin subspace ($d_s>1$) and the relevant $g_b$ operations act differently on different allowed spin components. As a result, some components are even and others are odd under $\mathbf{k}\rightarrow -\mathbf{k}$, giving a mixed-parity spin texture. Because such mixed parity requires at least two independent spin components, SST-4 can occur only in coplanar or noncoplanar textures. The corresponding symmetry conditions are summarized in Tab.~S4, and a representative $d_s=2$ tight-binding model is shown in Fig.~S4.\\
\indent Notably, the SST classification applies to collinear magnets as well: in the equilibrium SOC-free collinear limit only the even-parity SST-2 class survives, which contains the $s$-wave and the $d$-, $g$-, and $i$-wave altermagnetic textures\cite{alter1,alter2} as finer subclasses, whereas the SST-3 and SST-4 textures have no static collinear counterpart and are intrinsic to noncollinear order (see Appendix A, with a detailed derivation in Sec.~I of the SM\cite{sup}).\\
\begin{figure}[t]
\centering
\includegraphics[trim={0.0in 0.0in 0.0in 0.0in},clip,width=\linewidth]{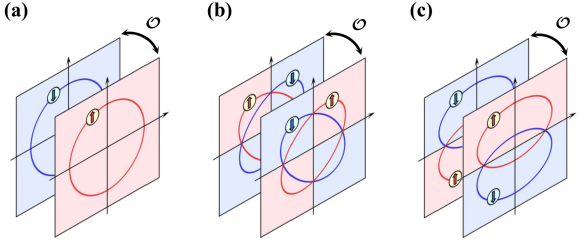}
\caption{Schematic illustration of HSP-1 in noncollinear magnets, depicting how local SST-1, SST-2, and SST-3 spin textures in (a)-(c) are constrained by the spin symmetry $\mathcal{O}$ to produce globally HSP.}\label{fig1}
\end{figure}
\begin{table*}[bhtp]
\caption{Classification of HSP in noncollinear magnets. The columns list, respectively, the symmetry constraining the dimensionality of the local spin polarization in reciprocal space, the resulting effective dimensionality, the possible types of local spin texture $\mathbf{S}^{\alpha}(\mathbf{k})$, and the key symmetry $\mathcal{O}$ responsible for the mutual compensation among local sectors, thereby generating HSP.}\label{Tab1}
\renewcommand{\arraystretch}{1.6}
\setlength{\tabcolsep}{5.5mm}{
\begin{tabular}{ccccc}
\hline  \hline
Prototype & Constraint & Dimensionality & $\mathbf{S}^{\alpha}(\mathbf{k})$ & Key symmetry $\mathcal{O}$     \\ \hline
HSP-1  & $[U_z(\phi)\Vert E\vert \tau]$ & collinear & SST-1; SST-2; SST-3 & $[T\Vert P\vert\tau]$; $[C_{2{\perp{z}}}\Vert E\vert\tau]$; $[TC_{2z}\Vert P\vert\tau]$    \\
HSP-2  & $[TC_{2z}\Vert P]$ & coplanar & SST-1; SST-2; SST-3; SST-4 & $[T\Vert P\vert\tau]$ ($[C_{2z}\Vert E\vert\tau]$)   \\
HSP-3  & --- & noncoplanar & SST-1; SST-2; SST-3; SST-4 & $[T\Vert P\vert\tau]$    \\
\hline \hline
\end{tabular}}
\end{table*}
\indent \emph{Hidden spin polarization in noncollinear magnets.}---To systematically investigate HSP in noncollinear magnets, we focus on a local sector exhibiting spin polarization (as described above) and identify a key spin operation, $\mathcal{O}$, that enforces compensation between local contributions, giving rise to the HSP phenomenon:
\begin{equation}\label{eq3}
\mathbf{S}^{\alpha}(\mathbf{k}) \xrightarrow{\,\,\, \mathcal{O} \,\,\,} -\mathbf{S}^{\beta}(\mathbf{k}),
\end{equation}
where $\mathbf{S}^{\alpha}(\mathbf{k})$ and $\mathbf{S}^{\beta}(\mathbf{k})$ represent the local spin polarizations of sectors $\alpha$ and $\beta$, respectively. Examining the action of spin-group operations on spin textures shows that HSP systems can be classified according to the spin dimensionality $d_s$ of the local spin polarization in reciprocal space. The resulting cases are discussed separately below and summarized in Tab.~1, with further details in Tab.~S5 of the SM\cite{sup}.\\
\indent \indent (i) For systems exhibiting the spin symmetry $[U_z(\phi)\Vert E\vert \tau]$, the local spin polarization in reciprocal space is constrained to be collinear, i.e., $d_s = 1$, and without loss of generality we take it along the $z$ axis. The global key symmetry $\mathcal{O}$ that enforces the hidden-spin relation between local sectors may then take the form $[T\Vert P\vert\tau]$, $[C_{2\perp z}\Vert E\vert\tau]$ (where $C_{2\perp z}$ denotes a twofold spin rotation about an axis perpendicular to $z$), or $[TC_{2z}\Vert P\vert\tau]$, each imposing the constraint
\[
\mathcal{O}\,\varepsilon(S_z,\mathbf{k})=\varepsilon(-S_z,\mathbf{k}),
\]
which defines the HSP-1 type. For a one-dimensional collinear spin texture, the local spin polarization can only belong to SST-1--SST-3, as illustrated schematically in Fig.~1.\\
\indent \indent (ii) For systems exhibiting the spin symmetry $[TC_{2z}\Vert P]$, the local spin polarization is constrained to be coplanar, i.e., $d_s = 2$, and we assume without loss of generality that it lies in the $x$--$y$ plane. In this case $\mathcal{O}$ may be either $[T\Vert P\vert\tau]$ or $[C_{2z}\Vert E\vert\tau]$, both of which require the band dispersion to satisfy
\[
\mathcal{O}\,\varepsilon(S_{x/y},\mathbf{k})=\varepsilon(-S_{x/y},\mathbf{k}),
\]
thereby defining the HSP-2 type. For the HSP-2 type, the local spin polarization may realize any of the four spin-texture classes, SST-1--SST-4. \\
\indent (iii) For systems lacking both $[U_z(\phi)\Vert E\vert \tau]$ and $[TC_{2z}\Vert P]$, the absence of additional constraints allows the local spin polarization to adopt a fully noncoplanar texture with $d_s = 3$ in reciprocal space. Symmetry analysis shows that the key symmetry $\mathcal{O}$ enforcing the mutual hiding of local spin polarizations is then uniquely $[T\Vert P\vert\tau]$, requiring $\mathcal{O}\,\varepsilon(S_{x/y/z},\mathbf{k})$ = $\varepsilon(-S_{x/y/z},\mathbf{k})$. A single symmetry is thus sufficient to govern the emergence of HSP-3, whose local spin polarization may likewise realize any of the four classes, SST-1--SST-4.\\
\indent In short, the classification thus involves two ingredients: the sector stabilizer symmetries specify what spin texture is hidden, namely its dimensionality $d_s$ and SST form, while $\mathcal{O}$ specifies how it is compensated globally. Absent such an operation, a globally observable nonrelativistic spin splitting is instead symmetry allowed. Throughout this work, the sectors are taken as the natural magnetic sublattices exchanged by $\mathcal{O}$, and the existence and category of HSP are independent of this partition (see Appendix B, with details in Sec.~II of the SM\cite{sup}).\\
\begin{figure}[t]
\centering
\includegraphics[trim={0.0in 0.0in 0.0in 0.0in},clip,width=\linewidth]{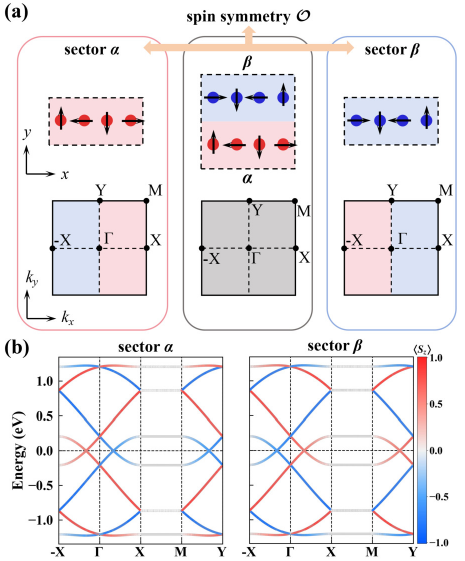}
\caption{Phenomenological model illustrating HSP. (a) Top: magnetic configuration, with red and blue spheres representing atoms in sectors $\alpha$ and $\beta$ and arrows indicating the spin ordering. Bottom: Brillouin zone of the spin distribution. Light red and light blue mark spin polarizations of opposite orientation in sectors $\alpha$ and $\beta$, whereas gray marks the regions where, constrained by the spin symmetry $\mathcal{O}$, the net spin polarization remains hidden. (b) Spin-resolved band structure, showing the contributions of the local sectors $\alpha$ and $\beta$ to $\langle S_z\rangle$.}\label{fig2}
\end{figure}
\indent \emph{Tight-binding model.}---Motivated by the above symmetry analysis, we construct a phenomenological tight-binding model describing electrons coupled to local magnetic moments to illustrate HSP. As shown in the top panel of Fig.~2(a), the Hamiltonian is written as 
\begin{equation}\label{eq:model}
\begin{aligned}
\mathcal{H} &= t\sum_{\langle i\alpha,j\alpha \rangle}c_{i\alpha}^\dagger c_{j\alpha} + \sum_{i,\sigma\sigma'} \mathbf{m}_i \cdot c_{i\sigma}^\dagger \boldsymbol{\sigma}_{\sigma\sigma'} c_{i\sigma'} + h.c.,
\end{aligned}
\end{equation}
where $c_{i\alpha(\beta)}^\dagger$ and $c_{i\alpha(\beta)}$ denote the creation and annihilation operators for electrons at site $i$ in sector $\alpha$ ($\beta$). The first term describes nearest-neighbor hoppings with amplitude $t = 0.5$, while the second term represents the exchange coupling between itinerant electrons and localized magnetic moments $\mathbf{m}_i$ with $|\mathbf{m}_i| = 0.5$, whose magnetic configuration is illustrated in Fig.~2(a). \\
\indent The model is built from two minimal $p$-wave sectors\cite{transport9} related by a key spin symmetry. Sector $\alpha$ [left panel of Fig.~2(a)] respects the spin symmetry $[C_{2z}\Vert E\vert\tau_1]$ with $\tau_1=(\tfrac{1}{2},0,0)$ together with the spin-only symmetry $[TC_{2z}\Vert E]$, which constrain the local spin texture $\mathbf{S}^{\alpha}(\mathbf{k})$ to be collinear and of odd-wave type, namely SST-3. The corresponding spin texture and spin-resolved band structure [Figs.~2(a) and 2(b)] exhibit alternating spin polarization along the $-X$--$\Gamma$--$X$ path, characteristic of a $p$-wave magnet. Sector $\beta$ is generated from sector $\alpha$ by the key spin symmetry $\mathcal{O}=[C_{2xy}\Vert E\vert\tau_2]$ with $\tau_2=(0,\tfrac{1}{2},0)$, so that it inherits the same local symmetry character while its spin polarization transforms as $S_z^{\alpha}(\mathbf{k}) \rightarrow -S_z^{\beta}(\mathbf{k})$. The two sectors therefore carry opposite local spin polarizations, whose contributions compensate in the full lattice and yield a global HSP [Figs.~2(a) and 2(b)], providing a concrete realization of the HSP-1 type. More generally, the key symmetry need not be limited to $\mathcal{O}=[C_{2xy}\Vert E\vert\tau_2]$: symmetries such as $[TC_{2z}\Vert P]$ or $[T\Vert P\vert\tau_2]$ can play an equivalent role in enforcing the hidden-spin relation between local sectors.\\
\begin{figure}[t]
\centering
\includegraphics[trim={0.0in 0.0in 0.0in 0.0in},clip,width=\linewidth]{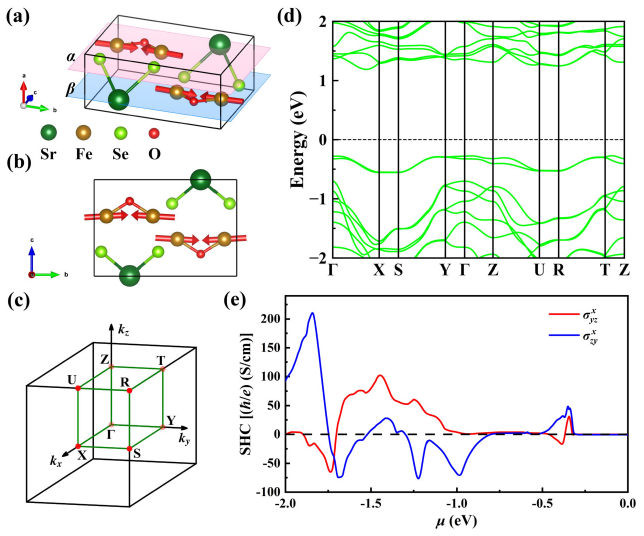}
\caption{(a, b) Crystal structure and magnetic configuration of SrFe$_2$Se$_2$O, with light-red and light-blue regions indicating the local $\alpha$ and $\beta$ sectors. (c) Three-dimensional Brillouin zone. (d) Nonrelativistic band structure. The finite sector-resolved spin polarization is shown in Fig.~S6. (e) Nonrelativistic SHC illustrating spin transport in noncollinear magnets.}\label{fig3}
\end{figure}
\indent \emph{Realization in realistic materials.}---Based on the above symmetry constraints, and after excluding collinear magnets and materials with fractional atomic occupancies, we identify 790 noncollinear magnetic compounds among the 2,186 entries of the MAGNDATA database\cite{MAGNDATA1,MAGNDATA2}, 516 with coplanar and 274 with noncoplanar order (see Fig.~S5). Our symmetry criteria yield HSP in 209 coplanar (40.5\%) and 70 noncoplanar (25.6\%) materials: the coplanar compounds are distributed among HSP-1, HSP-2, and HSP-3 with 122, 7, and 80 entries, and the noncoplanar ones among HSP-1 and HSP-3 with 11 and 59 entries, respectively (complete list and classification in Tab.~S6\cite{sup}). HSP is thus predominantly of the HSP-1 and HSP-3 types, reflecting that most local sectors break the $[TC_{2z}\Vert P]$ symmetry, which governs coplanar spin textures.\\
\indent As an illustrative example, we consider the noncollinear magnet SrFe$_2$Se$_2$O\cite{SrFe2Se2O_1,SrFe2Se2O_2} (see the SM for computational details\cite{sup}), a spin-ladder compound with long-range antiferromagnetic order in which oxygen mediates antiferromagnetic superexchange along the ladder rungs and selenium ferromagnetic coupling along the ladder direction. At 10 K the ordered magnetic moments point toward the center of the ladder and deviate slightly from perfect collinearity, giving a coplanar configuration in the $b$--$c$ plane that preserves both the spin symmetry $[TC_{2x}\Vert E]$ and the symmetry $[T\Vert P]$, and that divides the material into two sectors, $\alpha$ and $\beta$ [Figs.~3(a) and 3(b)]. Considering only the local $\alpha$ sector, $[TC_{2x}\Vert E]$ constrains the local spin texture to satisfy $\mathbf{S}^{\alpha}(\mathbf{k}) = (-S_x(-\mathbf{k}),\, S_y(-\mathbf{k}),\, S_z(-\mathbf{k}))$, so that $S_x(\mathbf{k})$ and $S_z(\mathbf{k})$ are odd and even functions of momentum, respectively, as shown in Figs.~S6(a) and S6(b). This corresponds to the SST-4 type, with locally mixed spin polarization in momentum space. The $[T\Vert P]$ symmetry linking the two sectors enforces hidden local spin polarizations, resulting in HSP of the HSP-3 type: no net spin polarization appears in the global band structure of Fig.~3(d), whereas the sector-resolved bands carry finite local spin expectation values [sector $\alpha$; Figs.~S6(a) and S6(b)]. Additional examples of materials exhibiting distinct local spin polarization are provided in Sec.~III of the SM\cite{sup}, including the HSP-1 type materials USb\cite{USb_1,USb_2,USb_3, USb_4}, Sr$_2$Mn$_3$Sb$_2$O$_2$\cite{Sr2Mn3Pn2O2_1,Sr2Mn3Pn2O2_2,Sr2Mn3Pn2O2_3}, PrFeAsO\cite{PrFeAsO_1,PrFeAsO_2,PrFeAsO_3,PrFeAsO_4}, and the HSP-3 type material GdMn$_2$Si$_2$\cite{GdMn2Si2_1,GdMn2Si2_2,GdMn2Si2_3}.\\
\indent Previous studies on HSP in nonmagnetic and collinear magnetic systems have demonstrated that such systems can exhibit spin-dependent responses\cite{spin_valve,Magnetoelectric,shc}. Extending this perspective, symmetry analysis indicates that noncollinear magnets with HSP can likewise host a variety of spin-dependent phenomena. As a representative example, we consider the intrinsic spin Hall conductivity (SHC)~\cite{shc1,shc2} in SrFe$_2$Se$_2$O. In the nonrelativistic regime, the spin Hall effect involves coupled dynamics in both real and spin spaces, and thus symmetry constraints must be imposed on both the spatial and spin indices of the SHC tensor (see Appendix C). For the coplanar magnetic phase of SrFe$_2$Se$_2$O this leaves only two independent symmetry-allowed components, $\sigma^{x}_{yz}$ and $\sigma^{x}_{zy}$. As shown in Fig.~3(e), the calculated nonrelativistic SHC fully conforms to this symmetry-imposed tensor form. In particular, $\sigma^{x}_{zy}$ is about $50~(\hbar/e)\,\mathrm{S/cm}$ near the Fermi level and can exceed $210~(\hbar/e)\,\mathrm{S/cm}$ at proper chemical potentials, indicating a sizable response that should be experimentally accessible. The SHC tensors for other representative noncollinear materials are summarized in Tab.~S7. \\
\indent Moreover, SrFe$_2$Se$_2$O exhibits a sector-resolved layer Hall effect in the absence of SOC: the two sectors related by the key symmetry  $\mathcal{O}$ carry finite but exactly opposite anomalous Hall conductivities and Berry-curvature distributions (see Fig.~S11), so that the local transverse responses remain finite while the total response vanishes. Unlike the SHC, which can also occur without any HSP, this compensated response is a direct, sector-sensitive signature of HSP (see Sec.~IV of the SM\cite{sup}). Beyond the $T$-even SHC and the SOC-independent layer Hall effect, HSP in noncollinear magnets can give rise to a much broader range of spin-dependent responses, summarized in Tab.~S9 and discussed in Appendix D.\\
\indent \emph{Summary and discussion.}---In summary, we establish a unified spin-group framework for HSP in noncollinear magnetic materials. Spin symmetries systematically constrain the nonrelativistic spin polarization, giving rise to four distinct local splitting forms which, combined with the dimensionality of the corresponding local spin textures and the symmetry relations between local sectors, naturally lead to three categories of HSP. We illustrate these categories using tight-binding models and representative materials, and a survey of the MAGNDATA database identifies 133, 7, and 139 candidate noncollinear magnets hosting HSP-1, HSP-2, and HSP-3, respectively. Our symmetry analysis and first-principles calculations further show that many of these materials exhibit nonzero spin-related response tensors, including an SOC-independent, sector-resolved layer Hall effect. These results provide a general symmetry-based understanding of HSP in noncollinear magnets and highlight their promise as a broad platform for spin-dependent functionalities.\\
\indent Because our classification and response analysis are formulated entirely in terms of spin groups, the results reported here should be understood as applying to the nonrelativistic, SOC-free limit. This point is particularly important for real materials, where SOC is generally present and may further modify the band splitting, spin textures, and response tensors. In this context, the present framework provides a useful symmetry benchmark for disentangling effects that are already enforced by noncollinear magnetic symmetry from those that are induced or enhanced by SOC. More broadly, our work places HSP in noncollinear magnets on a firm symmetry footing and identifies these systems as a promising platform for locally spin-polarized yet globally compensated electronic structures, with potential implications for low-dissipation spin transport, unconventional spin Hall phenomena, and symmetry-controlled spintronic functionalities.\\
\indent \emph{Acknowledgments.}---This work is supported by the National Natural Science Foundation of China (Grant No.~12574070 and No.~12504223), the Postgraduate Scientific Research Innovation Project of Hunan Province (CX20240616), the China Postdoctoral Science Foundation (Grants No.~2025M773383 and No.~GZC20252231), the China Postdoctoral Science Foundation-Hunan Joint Support Program (Grant No.~2025T002HN).  
\bibliography{references}
\onecolumngrid
\vspace{1em}
\begin{center}
	\textbf{\large End Matter}
\end{center}
\vspace{1em}
\twocolumngrid
\textit{Appendix A: Collinear limit of the SST classification.}---The SST classification rests only on the spin dimensionality $d_s$ fixed by the $g_a$-type operations and on the momentum parity fixed by the $g_b$-type operations, and therefore applies to collinear as well as noncollinear magnets in the nonrelativistic limit\cite{even_wave,odd_wave1, odd_wave2,odd_wave3}. The SST label specifies which spin components can be finite, whether they survive Brillouin-zone integration, and which spin-dependent response tensors they permit. Detailed discussion on physical significance of the SST classification is provided in Sec.~I of the SM\cite{sup}. \\ 
\indent The equilibrium SOC-free collinear limit is, however, strongly restricted. Every such magnet possesses the spin-preserving pseudo-time-reversal symmetry $[TC_{2\perp}\Vert E]$, with $C_{2\perp}$ a twofold spin rotation about an axis perpendicular to the common magnetic axis\cite{alter1,sgroup3}. Together with the collinear constraint $d_s=1$, this $g_b$-type symmetry forces the spin texture to be even in momentum, so that only the one-component SST-2 class survives. The ferromagnetic-like $s$-wave texture and the $d$-, $g$-, and $i$-wave altermagnetic textures\cite{alter1} are finer subclasses within this even-parity collinear class, distinguished by crystalline harmonics and symmetry-enforced nodal structures that are not encoded in the SST label itself. By contrast, the odd-parity SST-3 and, in particular, the mixed-parity SST-4 textures have no static collinear counterpart and are intrinsic to noncollinear order. A detailed derivation is given in Sec.~I of the SM\cite{sup}. 

\textit{Appendix B: Sector partition and symmetry criterion for HSP.}---The classification of HSP involves two logically distinct ingredients. The stabilizer symmetries of an individual sector determine the dimensionality $d_s$ and the SST form of the sector-resolved spin polarization, specifying what type of spin texture is hidden, whereas the sector-exchanging operation $\mathcal{O}$, acting at the same momentum, determines how the finite local textures compensate in the full crystal. \\
\indent The local sectors are taken throughout as the natural magnetic sublattices exchanged by $\mathcal{O}$, following the canonical convention of collinear HSP studies\cite{afm_hsp2}: each sector forms a spatially localized magnetic sublattice carrying a finite local spin polarization, while the total polarization vanishes by symmetry. Symmetry-equivalent sector choices or relabelings merely rotate the same texture and leave the SST assignment unchanged, and the existence and category of HSP are fixed by the full SSG, independent of the sector partition. The sector definition is provided in Sec.~II of the SM\cite{sup}. \\
\indent Conversely, if a finite local spin texture is symmetry allowed but the SSG contains no compensation operation $\mathcal{O}$ reversing all allowed spin components at the same momentum, the cancellation between sectors is not enforced and a globally observable nonrelativistic spin splitting becomes symmetry allowed. This criterion, whose admissible operations depend on $d_s$ as summarized in Tab.~1, generalizes the space--time-inversion and translation--spin-rotation conditions of collinear magnets\cite{afm_hsp2} and separates HSP magnets from magnets with global nonrelativistic spin splitting. In particular, the hidden altermagnetic spin splitting recently reported in collinear antiferromagnets\cite{afm_hsp4} corresponds to the collinear limit of HSP-1 with an SST-2 local texture, in which the compensation operation $[C_{2\perp z}\Vert E\vert\tau]$ reduces to the familiar spin-translation symmetry $[C_{2}\Vert \bm{t}]$. The general criterion, the relation to the collinear classification, and the role of noncollinearity are detailed in Sec.~II of the SM\cite{sup}. 

\textit{Appendix C: Nonrelativistic spin Hall conductivity.}---The intrinsic SHC introduced in the main text is evaluated as~\cite{shc1,shc2,shc3}
\begin{equation}
\begin{aligned}
\sigma_{\alpha\beta}^{\gamma} & =\frac{e}{\hbar}\sum_{n}\int_{BZ}\frac{d^{3}\textbf{k}}{(2\pi)^{3}}f_{n}(\textbf{k})\Omega_{n,\alpha\beta}^{\gamma}(\textbf{k}), \\
\Omega_{n,\alpha\beta}^{\gamma}(\textbf{k}) & = 2i\hbar^{2}\sum_{m\neq n}\frac{\langle u_{n}(\textbf{k})|\hat{J}_{\alpha}^{\gamma}|u_{m}(\textbf{k})\rangle\langle u_{m}(\textbf{k})|\hat{v}_{\beta}|u_{n}(\textbf{k})\rangle}{\left(E_{n}(\textbf{k})-E_{m}(\textbf{k})\right)^{2}},
\end{aligned}
\end{equation}
where $\Omega_{n,\alpha\beta}^{\gamma}(\textbf{k})$ denotes the spin Berry curvature and $f_{n}(\textbf{k})$ is the Fermi--Dirac occupation. The SHC tensor $\sigma_{\alpha\beta}^{\gamma}$ ($\alpha,\beta,\gamma = x,y,z$) characterizes the spin current $\hat{J}_{\alpha}^{\gamma} = \frac{1}{2}\{\hat{v}_{\alpha},\hat{s}_{\gamma}\}$ generated by an electric field $\bm{E}$ via $\hat{J}_{\alpha}^{s_{\gamma}} = \sigma_{\alpha\beta}^{\gamma} \bm{E}_{\beta}$, where the current flows along $\alpha$ with spin polarization along $\gamma$. Under a spin point group operation $[U_s \Vert R_g]$, the $T$-even SHC transforms as~\cite{TensorSymmetry}
\begin{equation}
[U_s \Vert R_g]\sigma_{\alpha\beta}^{\gamma} = \det|U_s| \sum_{\gamma' mn} U_{\gamma\gamma'} R_{\alpha m} R_{\beta n} \sigma_{mn}^{\gamma'},
\end{equation}
which yields the symmetry-allowed tensor components discussed in the main text and summarized in Tab.~S7. For the coplanar magnetic phase of SrFe$_2$Se$_2$O, the spin-only antiunitary symmetry $[TC_{2x}\Vert E]$ permits only SHC components with spin index $x$, namely $\sigma^{x}_{\alpha\beta}$. Combined with the operations of the nontrivial spin point group $^{1}m^{2z}m^{2y}m$, generated by $[E\Vert M_x]$, $[C_{2z}\Vert M_y]$, and $[C_{2y}\Vert M_z]$, this further restricts the SHC tensor to just two independent components: $\sigma^{x}_{yz}$ and $\sigma^{x}_{zy}$. 

\textit{Appendix D: Other spin-dependent response tensors.}---Table~S9 collects the spin-dependent response tensors permitted in the HSP noncollinear magnets identified above. They comprise magnetoelectric effects, magnetic susceptibility, and toroidal moments, together with antisymmetric components of transport tensors such as the Hall effect, thermal Hall effect, and Faraday rotation. Which components survive is dictated by the underlying spin symmetries, and antisymmetric contributions in particular can remain finite even when the global spin polarization vanishes. HSP magnets therefore combine a vanishing net spin with robust and diverse spin-dependent responses, providing a symmetry-based route toward spintronic functionalities in noncollinear magnetic materials.
\end{document}